\documentclass[letterpaper,twocolumn,10pt]{article}
\usepackage{usenix2025_SOUPS}

\usepackage{tikz}
\usepackage{amsmath}
\usepackage{dirtytalk}
\usepackage{balance}
\usepackage{graphics} 
\usepackage{booktabs}
\usepackage{xcolor,colortbl}
\usepackage{comment}
\usepackage{booktabs}

\usepackage{tabularx}
\usepackage{tipa}
\usepackage{enumitem}
\usepackage{url}

\usepackage{breakurl}
\usepackage{enumitem}
\usepackage{soul}
\usepackage{xspace}
\usepackage{microtype}
\usepackage{balance}
\usepackage{natbib}
\usepackage{tcolorbox}

\usepackage{footmisc}

\usepackage{etoolbox}
\patchcmd{\quote}{\rightmargin}{\leftmargin 0.5em \rightmargin}{}{}

\AtBeginEnvironment{tcolorbox}{\small}

\begin{document}

\date{}

\title{\Large \bf ``We are not Future-ready'': Understanding AI Privacy Risks and Existing Mitigation Strategies from the Perspective of AI Developers in Europe}

\def\plainauthor{Alexandra Klymenko, Stephen Meisenbacher, Patrick Gage Kelley, Sai Teja Peddinti, Kurt Thomas, Florian Matthes}

\author{
{\rm Alexandra Klymenko$^*$}\\
Technical University of Munich
\and
{\rm Stephen Meisenbacher$^*$}\\
Technical University of Munich
\and
{\rm Patrick Gage Kelley}\\
Google
\and
{\rm Sai Teja Peddinti}\\
Google
\and
{\rm Kurt Thomas}\\
Google
\and
{\rm Florian Matthes}\\
Technical University of Munich
}  

\maketitle
\def\thefootnote{*}\footnotetext{These authors contributed equally to this work.}
\renewcommand{\thefootnote}{\arabic{footnote}}
\thecopyright

\begin{abstract}
The proliferation of AI has sparked privacy concerns related to training data, model interfaces, downstream applications, and more. We interviewed 25 AI developers based in Europe to understand which privacy threats they believe pose the greatest risk to users, developers, and businesses and what protective strategies, if any, would help to mitigate them. We find that there is little consensus among AI developers on the relative ranking of privacy risks. These differences stem from salient reasoning patterns that often relate to human rather than purely technical factors. Furthermore, while AI developers are aware of proposed mitigation strategies for addressing these risks, they reported minimal real-world adoption. Our findings highlight both gaps and opportunities for empowering AI developers to better address privacy risks in AI.
\end{abstract}

\section{Introduction}
The recent emergence of widely accessible, general-purpose AI systems, including ChatGPT, Gemini, Claude, and Stable Diffusion, has sparked a renewed concern about the privacy risks posed by AI. These risks include models requiring immense amounts of training documents or images~\cite{jain2016big,jang-etal-2023-knowledge}, models memorizing and reproducing sensitive data~\cite{236216,274574}, model operators logging interactions with potentially sensitive data~\cite{khowaja2024chatgpt,bsi2024ai}, and potentially harmful applications of models~\cite{council2023ghost,golda2024privacy}. Along with studying known privacy risks, researchers have also explored mitigations via privacy-enhancing technologies (PETs), yet these investigations often reveal the limitations of such techniques, including balancing the privacy-utility trade-off \cite{golda2024privacy} and justifying computationally expensive methods \cite{vassilev2024adversarial}.

In response to the growing list of privacy risks posed by AI, regulators and technologists have worked to systematize known risks and gain perspectives on user concerns about AI. Within the European Union, regulators have codified the risks of AI systems in the upcoming AI Act\footnote{\url{https://eur-lex.europa.eu/eli/reg/2024/1689/oj}}. Researchers have also come up with their own systematizations of AI risks~\cite{shahriar2023survey, bsi2024ai, golda2024privacy}, which can be viewed as a subset of all known AI risks. In addition to enumerating AI privacy risks, these works also recognize the increasing importance of privacy legislation \cite{shahriar2023survey} and effective risk management strategies \cite{golda2024privacy}. Finally, studies have also explored user sentiment towards AI and privacy, where users’ fears of a decrease in privacy ranks amongst the top two concerns with AI ~\cite{kelley2023there}. Missing from these perspectives, however, are the views of developers.

In this work, we explore the current privacy risks of AI, their relative importance, and the efficacy of existing mitigation strategies from the perspective of 25 AI developers in Europe. These participants span the AI development stack, including research, data collection, model training and fine-tuning, and deployment or integration into enterprise systems (termed AI developers, for ease of reference). Through semi-structured interviews, we address three research questions: 

\vspace{5pt}
\noindent\textbf{RQ1:} What privacy risks do AI developers believe pose the largest risk to the users, developers, and businesses?\\
\textbf{RQ2:} What mitigation strategies for AI privacy risks have AI developers considered or deployed?\\
\textbf{RQ3:} What factors do AI developers think will influence the future of privacy in AI?
\vspace{5pt}

Our findings show that there is little consensus among AI developers on which known AI privacy risks are the most urgent to address. However, developers focused on three high-level categories more than any other: \textit{data management} (e.g., storage of model training data), \textit{data memorization and leakage} (e.g., large language models unintentionally leaking training text), and \textit{misuse via harmful applications} (e.g., deepfakes). Our discussion with developers reveals a variety of factors behind this divergence in risk ranking, including the recency of materialized privacy harms, the user or enterprise context, and ethical factors, among others. In addition, we observe that although general awareness of currently available mitigation strategies among developers is high, widespread adoption of mitigations is minimal beyond select methods, such as anonymization. Overall, developers expressed that they lack sufficient readiness to mitigate future privacy risks, citing reasons such as unpredictability in user engagement with models and lack of clarity coming from regulations. We synthesize these concerns, along with opportunities shared by developers, to understand how we as a community can work towards more privacy-conscious AI development.

We conclude that researchers should strive for greater clarity on how AI privacy risks are understood in practice, and advocate for a greater focus on building awareness of such risks, particularly to reach non-technical audiences such as end users or regulators. We also see that more effort is needed from the research community to make privacy risk mitigation strategies accessible to AI developers, so that advancements in PETs can be translated into practice. To summarize, we uncover the blueprints for an ecosystem in which all parties -- AI developers, end users, researchers, and
regulators -- must work in parallel towards a privacy-friendly future of AI.

\section{Related Work}
\textbf{Enumerating AI privacy risks and mitigations.}
In recent years, several studies have worked towards systematizing privacy risks in AI \cite{council2023ghost, curzon2021privacy, shahriar2023survey, yan2024protecting, bsi2024ai, golda2024privacy, rahman2023security, 10.1145/3613904.3642116, meisenbacher2024privacy}. In particular, these works tackle privacy risks from a largely technical perspective, identifying where privacy vulnerabilities may occur in AI systems. As each work presents its own collection of risks, the overall field is disjoint, with no unifying taxonomy. In addition, these works all represent an academic perspective, with no external validation from AI developers (e.g., readiness to mitigate such risks), adding to the lack of practicality.

Recent works have also surveyed existing techniques to mitigate privacy risks in AI systems \cite{curzon2021privacy, shahriar2023survey, bsi2024ai}. Mitigations presented in these works are also often comprehensive yet disjoint, and to the best of the authors' knowledge, none of these works seek to gain practical perspectives on the efficacy of such mitigations. We use these frameworks of risks and mitigations as the foundation for our interview guide, which features a ranking exercise of five high-level privacy risk categories and a graphic of all identified mitigation strategies.

\textbf{User and developer perspectives on AI privacy.}
On the user side, studies of concerns regarding the transformative nature of AI have been recently conducted with a particular focus on privacy \cite{kelley2023there}, finding that privacy ranks as the second highest concern following job loss. In this study, Kelley et al. \cite{kelley2023there} observe that nearly half of the roughly ten thousand survey respondents expect privacy to decrease over the next 10 years due to AI, citing reasons of model vulnerabilities, surveillance, unchecked data collection, and others. Such findings highlight the privacy implications of a rapidly developing landscape where AI has become increasingly ubiquitous.

On the practitioner side, previous works have focused on software developers and their privacy perceptions, for example, from the perspective of requirements engineering \cite{10.1007/978-3-030-44429-7_8} or privacy information in coding tasks \cite{maprogrammer}. Hadar et al. \cite{hadar2018privacy} conduct interviews with 27 developers and highlight the complex web of cognitive, organizational, and behavioral factors that play a role in privacy decision-making. Others explore the intersection of users and developers, finding that developers often may not align on what user privacy entails \cite{SENARATH20181845}.

Prybylo et al. \cite{prybylo2024evaluating} conclude that many software teams, regardless of region, are not entirely familiar with privacy concepts such as anonymization, and therefore they must rely on self-teaching and forums for privacy-related questions for their work. Interestingly, though, more than half of the 198 respondents in the study reported using some kind of PETs in their work, pointing to the increasing need for practical privacy protection in software systems.  Boenisch et al. \cite{10.1145/3473856.3473869} reach similar conclusions from a survey with ML practitioners, finding that awareness of security and privacy risks is low, and that privacy protection methods are less well-known than traditional security protections. These studies, however, do not focus specifically on AI developers, and the survey format precludes in-depth discussions on \textit{why} developers may not feel equipped to handle privacy risks. To fill this gap, we focus on developer \textit{awareness} and \textit{perception} of privacy risks in AI. 

In studying development teams building AI products, Lee et al. \citep{lee2024don} echo many of the same findings regarding unawareness of privacy threats in AI, leading to observations that many developers feel ill-equipped to perform privacy work. Lee et al. conclude by emphasizing the need to improve practitioner awareness of AI-specific privacy risks, and, correspondingly, to bolster their ability to address privacy harms. In our work, we maintain a focus on practitioner perspectives yet shift the target audience from various practitioner roles (designers, product managers, engineers) to exclusively technical practitioners developing, deploying, or integrating AI systems, thereby removing the bias of one business angle. While Lee et al. concentrate on processes, methods, motivators, and inhibitors in privacy work, we focus on awareness and perception of \textit{specific} privacy risks in AI and their mitigations. We build on the focus of Lee et al. on North American companies by exploring the European perspective, where data protection regulations play a significant role alongside impending AI regulations spearheaded by the European Union.

\section{Methodology}
We conducted semi-structured interviews with 25 AI developers in Europe to investigate their perception of AI systems'
privacy risks and potential mitigations. 
An AI system can be any of the General-Purpose AI Systems (GPAIS), as defined in an early draft of the AI Act\footnote{The final version of the EU AI Act specifies a different, more narrow definition of an AI system.}:

\begin{quote}
    \textit{General Purpose AI System (GPAIS): an AI system that is \say{able to perform generally applicable functions such as image/speech recognition, audio/video generation, pattern detection, question answering, translation, etc.} \cite{ai-act-draft}, and is able to have multiple intended and unintended purposes. Within the scope of GPAIS, we investigate Large Language Models, Diffusion Models, and Multimodal Models.}
\end{quote}

In the remainder of this work, we use the terms \textit{AI} and \textit{AI System} for simplification.

\subsection{Identifying Privacy Threats \& Mitigations}
In order to guide the discussion on AI risks and mitigations, we first conducted a literature review to build an overview of existing surveys and taxonomies. The main goal of this review was to collect a set of well-known privacy risks and mitigations to be presented during the interviews, and to systematize dozens of risks under various terminologies in a manner that is practical and comprehensible to participants within the scope of a single interview. The aim of this review was not to create a fully comprehensive privacy risk and mitigation taxonomy, but rather to establish an understandable and accessible basis for investigation in our interview study.

Our search yielded 4 journal papers, 3 conference papers, 4 whitepapers, and 2 preprints. We identified these resources using a methodology proposed by Kitchenham \cite{kitchenham2015evidence}, with the following search string (for titles only):
\begin{center}
    \scriptsize
    \textit{("privacy" OR "private") AND ("risks" OR "risk" OR "harms" OR "harm" OR "threats" OR "threat" OR "concerns" OR "concern" OR "dangers" OR "danger" OR "protect*" OR "mitigat*") AND ("AI" OR "Artificial Intelligence" OR "GPAIS" OR "General Purpose AI" OR "General Purpose Artificial Intelligence" OR "Language Model" OR "LLMs" OR "Generative AI" OR "Diffusion Model" OR "Multimodal Model")}
\end{center}

We utilized this search string in Google Scholar, which yielded 158 results. We reduced this to 148 documents by restricting our analysis to only papers from 2015 onwards. Next, we screened each document to remove results that: (1) were not accessible via public or institutional login, or (2) were specific case studies, legal opinions, or implementation proposals. We opted for this second criterion in order to emphasize papers of the \textit{survey} nature. Our screening isolated 6 relevant survey papers, from which we performed a forward/backward search and found 7 other relevant works, including gray literature sources \cite{garousi2019guidelines}, leading to the final set of \textbf{13} survey papers~\cite{council2023ghost, rigaki2023survey, curzon2021privacy, shahriar2023survey, vassilev2024adversarial, yan2024protecting, bsi2024ai, biggio2018wild, smith2023identifying, golda2024privacy, rahman2023security, mcgraw2024architectural, YAO2024100211}.

From these primary sources, our team worked collaboratively and iteratively to collate individual privacy risks into 5 high-level categories, as shown in Appendix \ref{sec:appendix:risks}. Similarly, we reviewed the sources for proposed mitigations, resulting in another 5 high-level categories, as shown in Appendix \ref{sec:appendix_mitigations}.

\subsection{Interview Design}
Below, we introduce the structure of the conducted interviews. The full interview guide can be found in Appendix \ref{sec:interview_guide}.

\textbf{Defining AI to participants.}
We began our interview with a discussion of the terminology \textit{General-Purpose AI System}, to align all participants on the interview context. 





\textbf{Background and experience with AI.}
We asked with which types of models from our GPAIS definition the participant had the most experience, and inquired about the centrality of AI in their work. This helped to establish the context in which we could analyze all ensuing responses. 

\textbf{Privacy risk awareness and perception (RQ1).}
This section allowed the participants to speak openly about what they believe are privacy risks in AI. We then asked about the role of AI in \textit{creating new risks} or \textit{exacerbating existing ones}. We first let participants reflect on what they believed privacy risks in AI to be, in order to gain an unbiased perspective before presenting our risk categories. We then invited the participant to take part in an interactive \textit{ranking exercise}, where the task was to rank our five risk categories (Table \ref{tab:risks} of Appendix \ref{sec:appendix:risks}) based on the \textit{urgency to address}, and provide a reasoning for the chosen ranking. Each category was accompanied by a brief description. The order as given in Table \ref{tab:risks} mirrors the ordering as presented in the ranking exercise; the order was determined by the AI lifecycle stage. To allow for free and open discussion, we did not restrict the risk context, for example, from the user, developer, or business perspective.

\textbf{Mitigation strategy awareness, perception, and readiness (RQ2).}
We then switched to mitigations, once again openly asking for any mitigations with which the participant was familiar, in order to understand their general awareness of mitigation strategies. We then presented a structured list of mitigations identified in the literature (see Appendix, Table \ref{tab:mitigations}), inviting comments and discussion on the presented strategies. Finally, we asked whether any of the presented mitigations the participant believed to be particularly effective (or not), thus not only assessing familiarity but also perception. As with the presented risk categories, each mitigation strategy was presented alongside a brief definition, in order to facilitate a general understanding for the interview participant.


We also gauged the participant's perception of current readiness to mitigate privacy risks, both personally and generally speaking. We first asked, given current mitigation strategies, which privacy risks the participant believes to be most addressable, and which are currently still challenging or difficult to mitigate. This question was then repeated, but now asking the participant to assess their personal readiness. As a last question, we asked the participant to express what would help to increase their readiness to mitigate privacy risks.

As an additional exercise, we asked each of our participants to partake in an offline survey following the interview. In this survey, we asked the participants to reflect on their familiarity and experiences with each of the 18 presented mitigation strategies. Specifically, we asked for a self-assessed selection of one of the following options: \textit{never heard of it}, \textit{heard of it}, \textit{considered using it}, \textit{used/using it}.

\textbf{Factors influencing the future of privacy in AI (RQ3).}
Following both of the above exercises, we asked open-ended questions for the participants to reflect on the factors that might influence the future of privacy in AI.

\textbf{Pilot interviews.} After creating the final draft of the interview guide, we conducted two pilot interviews, focusing on identifying any ambiguities in the questions. As a result, no major changes were made to the guide, only minor ordering and stylistic edits. 
The results of the two pilot interviews (I1, I2) were included in the final analysis and results.

\subsection{Study Protocol}
\noindent\textbf{Recruitment.}
We sought to garner a wide European geographic diversity of participants while maintaining our target audience of technical AI developers. We recruited via snowball sampling through two sources: personal contacts and AI developers via LinkedIn. For the latter, we searched for relevant keywords such as \say{AI Engineer}, \say{ML Engineer}, and \say{AI Consultant} and then screened promising profiles, particularly to confirm that the candidate fits our profile of a technical stakeholder working in AI (research, development, deployment, or integration). While we reached out to over 150 fitting LinkedIn profiles, we were able to establish formal email contact with 45. From these, we successfully conducted 17 interviews, in addition to 8 personal contacts or referrals.

\noindent\textbf{Interviews.}
We conducted a total of 25 interviews. All interviews were held in English. Two researchers were always present in the interviews. 
While one researcher was asking the questions, the other would be taking notes and generating potential follow-up questions. All interviews were planned for 60 minutes, and the average length of audio recording was 53 minutes (not including introduction and post-talks). Participants were not compensated for taking part in the study. The interviews were conducted from June to August 2024.

\noindent\textbf{Demographics.} 
Participants in our study spanned 12 European countries\footnote{One participant works for a European company, but is located in India.}. Most worked directly on AI systems (14), while the remainder worked on integrating AI features (11). Years of experience working with AI ranged from 1-3 (1), 3-5 (5), 5-10 (5), 10-20 (9), and 20+ (5). Education attainment was generally high: all participants had either a bachelor's degree (1), master's degree (11), or doctorate degree (13). Regarding gender, 21 participants identified as male, 3 as female, and 1 preferred not to say (full details in Appendix Table~\ref{tab:interview_participants}).

\subsection{Data Processing and Analysis}
After transcribing each interview using Otter.ai\footnote{The quality of transcriptions was manually verified after each interview.}, we conducted a thematic content analysis \cite{braun2006using} on the interview data, working collaboratively with our research team. The two primary researchers began coding all transcripts by initially highlighting excerpts of interest. Then, in an iterative process, we grouped selected excerpts together into \textit{themes}, where we unified similar sentiments under a common message. As this was done collaboratively, some themes eventually were merged or deleted. We held regular meetings to synchronize the coding process, and to discuss any possible conflicts or suggestions for edits. In the end, we agreed upon a final set of themes, supported by feedback from our wider research group. Once no new themes were discovered during analysis, the interview study was concluded \cite{urquhart2013grounded,saunders2018saturation}.

With the set of codes, we conducted a \textit{synthesis} phase in which the goal was to construct a coherent narrative from the collection of codes. This was done in two steps, where the two main researchers jointly performed an initial synthesis, followed by a feedback round with all researchers, and then the implementation of feedback to form the final synthesis. This synthesized report forms the basis of our results.

\subsection{Ethics and Limitations}
Before we conducted interviews, we sent an email to potential candidates with the interview consent form, an intake survey with background information, and the full interview guide.
We shared the interview guide so that the participants felt prepared to answer our questions. The consent form clearly stated the terms of the interview, most importantly that the participation is voluntary, the interviews would remain confidential, with all PII pseudonymized, and that the participant consented to the recording, transcription, and publication of the results in a pseudonymized form. 
The participants were given the option to refrain from answering specific questions, as well as from sharing any company details or work practices. This study protocol was approved by the IRB of the Technical University of Munich, with approval \#2024-35-NM-BA.

We caution that our participants may not be representative of all European AI developers. Although we defined a clear candidate persona and performed profile screening, the selection process nevertheless followed a somewhat opportunistic approach, where a suitable candidate would immediately be invited once identified. This limitation is partially mitigated by the observed diversity of participants, from the perspective of educational background, industry domain, organization size, experience, and geographic location. We note, however, the relatively high education level of our participants, which could lead to a skewed perspective, as well as the clear geographical focus on Europe. Another limitation is the clear bias in gender distribution, with only 12\% women in participation, which lies beneath the current estimate of women in AI (29\%\footnote{\url{https://www.randstad.com/randstad-ai-equity/}}). A final limitation is inherent to our chosen method of semi-structured interviews, where willingness to participate and the resulting interview data are affected by social desirability and self-reporting bias. This bias could have potentially been amplified due to our decision to distribute the interview guide to each participant before the interview.

\section{Results}
We introduce our findings, supported by relevant interview quotes, indicated by the participant ID displayed in Table \ref{tab:interview_participants}. 

\subsection{Privacy Risks}
\label{sec:privacy_risks}

\subsubsection{Forces behind privacy tensions with AI}
To begin the discussions with our participants, we invited them to share their thoughts on whether and how privacy risks in modern AI systems are different from \say{general} privacy risks that can arise in any software system.

Participants acknowledged that many of the AI privacy risks in our risk categories are, in fact, not new when compared to privacy risks in other technical systems. Above all, it was seen that where there is data, there is risk, especially when there is some kind of query access to the data, e.g, model prompts (I2), as well as risks inherently stemming from black-box AI models. Especially in the European context, privacy has been a major topic since the introduction of GDPR.

However, there are a wide variety of reasons why developers perceive privacy risks to have been \textit{amplified} by recent advances in AI. While general privacy risks may be well known in popular media and among users, we take a more nuanced view with developers, focusing on specific technical privacy risks.  In addition to the general increase in awareness on the topic, we learned of several reasons perceived to be important in the rise of privacy concerns with AI.

\textbf{Increased Accessibility.}
AI developers perceived that although the risks of AI have existed for some time, these risks have been amplified by an increased accessibility to AI systems, including for non-technical users. Many participants (10/25) pointed to 
the trend that AI has been made much more accessible, giving examples of the ease of using pay-per-use, user-friendly interfaces or APIs.
Others also talked about the \say{human nature} of these interfaces, thus enabling users potentially to place false trust in AI systems. This was seen as potentially leading to a \textit{\say{lack of basic understanding}} (I22), contrasted to previously when \textit{\say{you needed a full team}} (I5) to develop, deploy, and even use AI models. Because of this shift in the way in which AI is provisioned, access to AI systems might be available to those unaware of their privacy implications, leading to uninformed usage and \say{reckless} behavior, something with which several participants were concerned (6/25). As one participant shared:
\begin{quote}
    \textit{\say{The biggest risk is that employees and also end users become more and more uncritical [about] what happens with their own data inside these large language models (...) because they want to have this new functionality.}} (I12) 
\end{quote}
Thus, increased accessibility to AI has exacerbated privacy issues, particularly due to a shift in the user base from research environments to the larger, generally non-technical public.

\textbf{Increased Capabilities.}
Many participants cite the considerably more advanced technical capabilities of modern AI systems as an amplifier of privacy risks (10/25). 
Older AI models also posed privacy risks, \textit{\say{but with the bigger models, you have more weights, more parameters, and I guess this increases the ability to memorize training data}} (I10). As a final piece to the puzzle, I23 points out that with such improvements, we can now leverage more data than before:
\begin{quote}
    \textit{\say{For instance, if I went back 10 years ago, I could have easily had a crawler and still do the same thing. But what would I do with that data, there was no algorithm or there was no model. So there was no incentive to do that.}} (I23)
\end{quote}
In this, we see a progression towards increased surface for privacy risks to transpire, rooted in the capabilities enabled by better and larger AI systems trained on increasingly larger data. Better AI systems also open the door to more capable malicious users, a point emphasized by a number of participants (4/25), as well as less concern about privacy in favor of functionality: \textit{\say{The benefits are so high in certain use cases that people are using it with less sensitivity}} (I7).

\textbf{Increased Use of Data.}
Advancements in AI training and deployment have allowed for the usage of larger amounts of data to train better models. The risks of the sheer size of such data in comparison to classic methods are made clear:
\begin{quote}
   \textit{\say{When we talk about classical ML training, you might be talking about a few thousand, tens of thousands of records. And here you're talking about so many terabytes of data. So you don't know what is actually going into there.}} (I11)
\end{quote}
An important extension emphasized by multiple AI developers (7/25) pertains not only to the size, but also to the \textit{value} of data. Specifically, data being sent by users to AI systems is both more sensitive and more valuable \textit{\say{because the data that you send to an LLM is, by definition, good training data}} (I7). Pointing to this fact, participants listed several privacy-related consequences, such as data leakage and profiling.

\begin{tcolorbox}[boxsep=3pt,left=1pt,right=1pt,top=1pt,bottom=1pt]
\textbf{Key Finding:} While many of the interviewed developers saw that some privacy risks are not new with AI, they pointed to the key \textit{amplification} factors of increased \textit{accessibility}, \textit{capabilities}, and \textit{use of data} that have been enabled by recent AI advances. 
\end{tcolorbox}

\begin{table}[t]
\centering
\footnotesize
\resizebox{\linewidth}{!}{
\begin{tabular}{l|ccccc|ccc}
\toprule
\multicolumn{1}{c|}{} & \multicolumn{5}{c|}{\textbf{No. of times ranked as:}} & \textbf{Median} & \textbf{Average} & \textbf{Standard} \\
\multicolumn{1}{c|}{} & \textbf{\#1} & \textbf{\#2} & \textbf{\#3} & \textbf{\#4} & \textbf{\#5} & \textbf{rank} & \textbf{rank} & \textbf{deviation} \\ \hline
\textbf{Misuse via Harmful Applications} & 7 & 6 & 4 & 3 & 5 & 2 & 2.72 & 1.51 \\
\textbf{Data Management} & 7 & 5 & 4 & 5 & 4 & 3 & 2.76 & 1.48 \\
\textbf{Data Memorization and Leakage} & 7 & 4 & 2 & 6 & 6 & 3 & 3.00 & 1.61 \\
\textbf{Unintended Downstream Usage} & 3 & 5 & 6 & 5 & 6 & 3 & 3.24 & 1.36 \\
\textbf{Adversarial Inference and   Inversion} & 1 & 5 & 9 & 6 & 4 & 3 & 3.28 & 1.10 \\
\bottomrule
\end{tabular}
}
\caption{AI developers' ranking of AI risks based on their potential harm and urgency to address (highest first).}
\label{tab:risk_ranking}
\end{table}

\subsubsection{Perceptions of AI privacy risks}

\textbf{Misuse via Harmful Applications.}
Many participants acknowledged that misuse is dangerous to privacy, often citing deepfakes, fake news, misinformation, propaganda, impersonation, and others. These types of threats were seen to be of great societal importance, especially in the modern day. In addition, such risks were tied to the idea of mental health, which can be perceived to be \say{privacy-adjacent}. In this light, the risk of misuse was seemingly the most \say{tangible} to many of the participants. They were able to give concrete examples of how this might affect people directly, such as through political deepfakes or family relative impersonation.

\begin{quote}
    \textit{\say{I find this whole news and misinformation and fake news, deep fakes, very problematic. And it's not just us technical people that are affected by this.}} (I10)
\end{quote}

While the discussions led to many insights regarding the ethical use of AI, these were considered outside the scope of this work focused on privacy risks. In essence, most of the participants pointed to the harms potentially resulting from the misuse, rather than the risk of misuse happening itself. In contrast to other discussed risks, where the harms described by the participants were largely confined to privacy harms, the concerns expressed under this risk category often extended beyond privacy to more general societal harms (e.g., the threat of misinformation and propaganda). This observation implies that the relatively high ranking of the risk of misuse may be entangled with broader concerns with AI, including privacy.

\textbf{Data Management.}
Many participants saw proper data management as a crucial first step in AI pipelines, stating that mitigation of risks here is the best mitigation for other risks such as data leakage and adversarial attacks. It was clear to many participants that the scale at which data is being collected necessitates strong data management practices. Thus, the risk lies in the fact that \textit{\say{if the baseline is problematic, then you're just making a big problem for the system itself}} (I6). Another participant highlights the danger of large-scale data collection without proper processes: \textit{\say{Companies nowadays, they just want to collect more and more data, just following the high-level understanding that more data is better.}} (I25)

Nevertheless, others expressed that data management might not be as important as other AI privacy risks because it is not specific to AI, cannot prevent errors in training procedures, is not relevant if models are trained on short-term stored (or streamed) data that is deleted post-training, and is a non-issue if the data itself was collected improperly. It was particularly stressed by I24 that \textit{\say{some of the what's captured in the model ends up being more important [than data management] because the model lives longer}} (I24), presenting an opposite perspective to data management being \textit{\say{at the core}} (I13).

\textbf{Data Memorization and Leakage.}
The concepts of data memorization and leakage were familiar to many participants, demonstrated by the repetition of popular examples from recent related papers. Acknowledging that memorization is an issue, participants hinted at the \say{garbage in, garbage out} principle, saying \textit{\say{don't put PII if you don't want them outputted. Simple as that.}} (I22). However, others saw the issue as more complex, expressing that it may be very difficult to determine if privacy has been compromised if we are not even sure what has been encoded in a model. Furthermore, controlling data leakage is \textit{\say{really very hard to control for developers}} (I17). Still, data leakage was not seen across the board as dangerous, as it might not happen often and \textit{\say{the difference between data leakage and hallucination... most of the times you can't really see that as an end user}} (I18).

\textbf{Unintended Downstream Usage (of Sensitive Prompts and User Metadata).}
The notion of unintended downstream usage attracted the least comments, yet it was seen as \textit{\say{specific to AI systems}} (I2). Placing sensitive information into prompts could pose privacy risks, and furthermore, carelessness in the processing of prompting can lead to important (company) information being leaked. Interestingly, also considered under this category was profiling via user metadata, as well as the uncertainty in many AI systems as to how the system responds to prompts, e.g., what the system prompt is.

\textbf{Adversarial Inference and Inversion.}
Adversarial inference and inversion attacks were at times tied to data memorization and leakage risks (4/25), pointing to their intertwined nature. In some cases (5/25), participants commented on the fact that such attacks are quite hard to execute and are not that common, and therefore are not as urgent, \textit{\say{not because it's less risky, but because at least nowadays, you need some skills to do that}} (I13). The same participant ranks this risk as the least urgent, saying that since most people have good intentions, we should focus on the risks that affect this majority, i.e., unintentional privacy risks. From another perspective, context is very important for adversarial privacy risks, and such risks might not be applicable in some settings:
\begin{quote}
    \textit{\say{If it is a public system, probably there will be a lot of adversarial attacks (...) But at least in my experience, for more controlled environments, it's not a big issue.}} (I16)
\end{quote}
Beyond context, there was potentially less urgency placed in adversarial risks due to a lack of clarity on what the risk entails, suggesting the importance of the \textit{tangibility} of privacy risks, which may impact their perceived urgency. 

\subsubsection{Ranking criteria for privacy risk importance}
In the ranking exercise of each interview, we asked the participant to rank the five privacy risk categories of Table \ref{tab:risks} in terms of importance, or rather, \textit{urgency to address}. We observed little unified consensus on risk importance (see Table \ref{tab:risk_ranking}), albeit with a clear top-3. We also asked the participants for their reasoning behind their ranking, which led to some salient \textit{reasoning patterns} behind perceived risk importance.

\begin{itemize}
    \itemsep 0em
    \item \textbf{Cause and Effect}: If a risk is the impetus for other risks, then it is perceived to be more important. Most notably, five participants believed that improper data management may realize other privacy risks.
    \item \textbf{Intent}: If a risk transpires \textit{unintentionally} (accidentally, without malicious intent), this was perceived by some to be worse than harms caused intentionally.
    \item \textbf{Difficulty}: Some participants stated that if a risk is technically more advanced, requiring a truly capable adversary, then this is perceived to be \textit{less} urgent. This is tied to the fact that such attacks may be less \textit{likely}.
    \item \textbf{Context is Key}: An important distinction is the enterprise (company) vs. individual (role) context, e.g., risks related to data management might be more relevant in the enterprise context. I25 asserts that \textit{role} is important: different levels of hierarchy have different priorities.
    \item \textbf{Existence of Solutions}: Sometimes, participants ranked privacy risks lower when they felt that solutions already existed to help mitigate them. Most notably, the risks of data management and data memorization and leakage were sometimes cited as less urgent for this reason.
    \item \textbf{Recent Privacy Incidents}: In select cases, the higher ranking of a risk was accompanied by the recounting of a recent incident, showing the linkage between risks and incidents in the minds of developers.
\end{itemize}

\begin{tcolorbox}[boxsep=3pt,left=1pt,right=1pt,top=1pt,bottom=1pt]
\textbf{Key Finding:} Beyond individual insights on AI privacy risks, we learn of a number of \textit{reasoning patterns} behind perceived risk prevalence, ranging from \textit{technical difficulty} to \textit{risk context}.
\end{tcolorbox}

\subsubsection{How addressable are these risks?}
We asked \textit{how addressable} the presented privacy risks were. In response, some participants were confident that certain privacy risks can be addressed given current mitigation strategies, saying that it is a matter of effort. Other participants, however, were less optimistic, pointing out that certain risks are quite unpredictable and, therefore, difficult to mitigate. 
For example, data management was often perceived positively:
\begin{quote}
    \textit{\say{Data management is the one we're most able to address, and that was also a factor in me putting it as the least urgent (...) because we do know how to do data governance.}} (I24)
\end{quote}
I25 echoed this sentiment, saying that with \textit{\say{clear policies, clear guidelines}} on both the technical and management level, mitigation of risks is manageable. Similarly, it was perceived that data memorization and leakage can be well addressed if companies make a choice to implement techniques such as Differential Privacy, and  \textit{\say{are not in a hurry to push out models}} (I23). One opinion states that many risks are mitigatable:
\begin{quote}
    \textit{\say{I think all the technological ones with a bit of effort, we are able to mitigate. Incorporating these technologies is just a matter of effort.}} (I13)
\end{quote}

Many of the risks believed to be \textit{less} mitigatable revolve around a common theme: \textit{that which cannot be easily predicted or controlled}, or rather, \textit{\say{any risk which is fuzzy}} (I5). Explicit examples of such risks, as given by the participants, include unpredictable prompts (I2, I10) and unpredictable users (I18, I24), lack of explainability (I7, I15, I17), and difficulty in detecting what PII is (I16). A particularly interesting aspect is the lack of clarity surrounding the legal requirements with respect to AI systems (I10, I13), which will be discussed in Section~\ref{sec:technical_legal_ethical}. Also of note is the fact that data management (I7, I15) and data memorization (I24) were also listed as least addressable, further highlighting the lack of consensus. 


\subsubsection{Technical, legal, and ethical risks}
\label{sec:technical_legal_ethical}
When asking generally about the relationship between the three aspects of privacy risks (technical, legal, and ethical), we received \say{orderings} of their relative importance.

Some participants (3/25) see the privacy problem foremost as a technical problem requiring technical solutions, as \textit{\say{first comes the technologies, the trends, the potential, and then the rights}} (I14). Other participants expressed that regulations lag behind technical advancements, and therefore, the technical nature is more important. 
As we only interviewed technical roles, these opinions can naturally be biased.

Focusing on ethical risks of AI was perceived by some participants to be of higher importance than legal ones (4/25), as they capture \textit{\say{concerns about people losing their rights more than the legal}} (I24). Nevertheless, they are all interlinked:
\begin{quote}
    \textit{\say{[T]he societal impact of it is, in the end, what matters. Technology I would see more as a means to get there. And the same thing for regulation, they should all make sure it has a positive effect on society.}} (I18)
\end{quote}
Some participants agreed that 
responsible AI should be reinforced in developers, such that they are \textit{\say{responsible for what they are doing (...) [since] it is for the common good}} (I4).

Finally, a number of participants (7/25) emphasized that ethical thinking and decision-making are more important than technical risks, because they guide the way in which technology is designed: \textit{\say{It needs to start with making a good ethical decision in the beginning.}} (I3). Moreover, in comparing technical risks versus societal risks, I15 expresses that it is hard to solve privacy risks, which are inherently human-oriented, with technical means, and I14 echoes this sentiment:
\begin{quote}
    \textit{\say{There are many factors that must be counted when solving technical problems, but they can be solved by reasoning, but (...) people are not problems you can solve.}} (I14)
\end{quote}
Regarding the potential harms of privacy risks, I23 says that technical risks may result in financial loss, \textit{\say{but if you start molding a society in a certain way, there is no coming back}}. This shows that some developers have extrapolated the implications of privacy risks well beyond technical vulnerabilities.

\begin{tcolorbox}[boxsep=3pt,left=1pt,right=1pt,top=1pt,bottom=1pt]
\textbf{Key Finding:} Developers perceive the technical, legal, and societal aspects of AI privacy risks to be intertwined, with particular concerns linking privacy risks to broader ethical considerations.
\end{tcolorbox}

\subsubsection{Other risks}

Some participants (7/25) also suggested risks that were not directly captured under our risk categorization. The reliance of some AI providers on \textit{third-party services}, including open-source libraries, was seen as potentially dangerous. 
Other participants brought up the risk of \textit{bias} as adjacent to privacy. 
Finally, the risk of \textit{not having established processes or standards} for developing or deploying AI systems was seen as a danger by some developers, as opposed to fields with standardized roles such as the medical or legal domains.

\subsection{Privacy Risk Mitigation Strategies}
\label{sec:mitigations}
\begin{figure}[t!]
    \centering
        \includegraphics[width=0.99\linewidth]{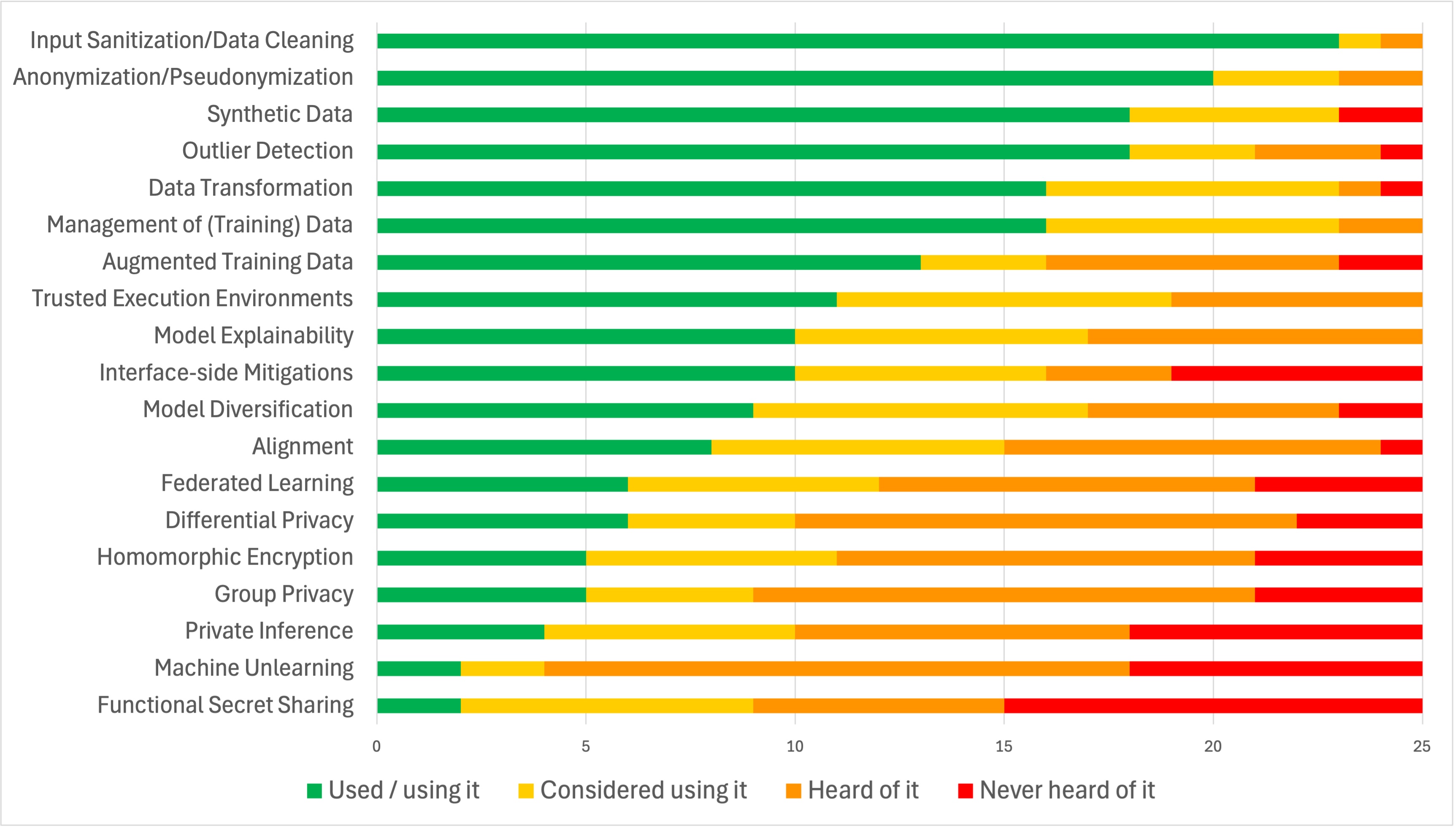}
    \caption{Participant self-ranking of their mitigation familiarity and experience in adopting these techniques.}
    \label{fig:mitigations_familiarity}
\end{figure}

We gained insights into the perceived effectiveness of mitigations as a whole, as well as individual perspectives on the strategies. To guide the discussion, we include in Figure~\ref{fig:mitigations_familiarity} the results of our supplemental survey asking participants about their level of familiarity with each of the 19 mitigations.

\subsubsection{Which mitigations are most (or least) effective?}
We wanted to know what AI developers thought were the most effective current mitigation strategies, whether it be the ones presented directly to them (Table \ref{tab:mitigations}) or otherwise. Instead of asking for comments on each mitigation strategy, we left the discussion open to the participant, and we asked instead for comments on the most familiar techniques, as well as those believed by the participant to be most or least effective. Generally speaking, there was no major consensus on which mitigation strategies are the \say{most effective} in mitigating privacy risks. When posed with this question, developers had varying responses, as well as \say{it depends} types of answers.

Many participants (7/25) stated explicitly that data sanitization techniques are the most effective in mitigating privacy risks in AI systems, aligning with the high amount of reported usage in Figure \ref{fig:mitigations_familiarity}. These developers saw it as straightforward techniques to remove PII, 
while others (3/25) saw general organization measures, including proper decision-making, as the best techniques for privacy risk mitigation: \textit{\say{the biggest impact you can make is reduce the number of bad decisions that someone in an organization makes}} (I3).

From a technical standpoint, smaller subsets of participants each supported different mitigation strategies as being very effective, from Differential Privacy (I1, I2), to Synthetic Data (I1, I5, I19, I20), to Federated Learning (I21). Additionally, some named techniques that we did not present, such as strong data policies (I16) or active monitoring systems (I18). Nevertheless, I13 points out that no single mitigation strategy will be a silver bullet: \textit{\say{in general, I think you need to use a combination of those. And it depends on the use case.}}

Notably, developers had fewer opinions on what the \say{least effective} mitigations might be, often not directly answering the question. Some isolated opinions included explainability (contradicting research, I21), organizational (hard to implement, I21), and Trusted Execution Environments (not practical at scale, I24). This could show that while developers may be generally familiar with mitigations, they are not able to judge their efficacy easily, or rather, to explain the downsides.


\subsubsection{Perceptions of individual mitigation strategies}
We present insights gained from participants on each of the 19 mitigation strategies, delineated by the five mitigation \textit{categories} (Table \ref{tab:mitigations}). For each mitigation, we list how many participants provided substantial comments on the technique.

\textbf{Organizational.}
Management of Training Data (4/25) was stressed as an important organizational process, especially with the GDPR mandate. However, it was noted that there exists a gap between research and practice, with a lack of good data management practices from research.

While Alignment (7/25) was seen by one to be quite important, most of the feedback focused on limitations, as well as it often not being a priority for organizations. The challenges of over-alignment and proper evaluation were also emphasized.

In a similar way, the discussion on Trusted Execution Environments (3/25) focused on whether they can be implemented in a cost-effective way, and the dependence on organizational infrastructure to determine when hardware might be secure.


\textbf{Data Preprocessing.}
While the strategy of Input Sanitization / Data Cleaning (11/25) attracted a great deal of attention, there was a split opinion on its effectiveness. While some championed it as \say{essential} and \say{key} (7/25), others emphasized that truly sanitizing data is more difficult than it seems, especially when considering the involvement of third parties or when dealing with highly unstructured data formats. 

A similar sentiment was observed in regard to Anonymization / Pseudonymization (6/25). Some participants used the language that such techniques are \say{common} and should be used \say{by default}. A number of participants (4/25), though, emphasized limiting factors in true anonymization, such as choosing the right parameters, high cost, and incorrect implementations leading to faulty anonymization. 

Outlier Detection (5/25) was perceived as directly related to privacy, and several participants regarded it as a support tool for other mitigations, such as Differential Privacy. 


\textbf{Data Augmentation.}
The number of supporters for Differential Privacy (9/25) was significant, with several alluding to it being the \textit{\say{gold standard}} (I10). These participants pointed to experiences in implementing it for AI applications, and I2 credits the maturity and availability of open-source resources, including a wide body of academic literature, as being helpful in facilitating adoption. At the same time, some developers mentioned challenges, such as in measuring privacy (I13), finding an acceptable privacy-utility trade-off (I10), and implementing it in already noisy domains (I7).

The notion of Group Privacy (3/25) attracted little attention, with the majority of the participants not being familiar with concepts such as k-anonymity. Interestingly, it was noted that such techniques are rarely used in production (I1), and two candidates (I10, I13) made a direct comparison to Differential Privacy, explaining its merits compared to Group Privacy.

Data Transformation methods (3/25) were met with skepticism, with participants pointing out that \textit{\say{you can partly obfuscate the data, but not fully}} (I22), and it is unclear whether these actually fulfill the purpose of privacy protection (I7). 

Although multiple participants commented on Homomorphic Encryption (7/25), the perceptions of this technology were overwhelmingly negative, with several (4/25) concerns about its unclear applicability to AI models. In addition, the high computational cost and slowness of Homomorphic Encryption were emphasized by I1, and I22 indicated a lack of standardization, making practical applicability difficult.

Functional Secret Sharing (1/25) received the least feedback out of \textit{any} mitigation strategy, and we noticed that the terminology itself was quite unclear to many participants. Only one participant (I10) was able to make a connection to a particular technique, i.e., Secure Multi-Party Computation.

In contrast, Synthetic Data (15/25) was touted for being \say{commonly} and \say{widely} used (I8, I17, I21) for training purposes, as well as helpful in privacy-critical scenarios, such as with medical data (I20, I22), or particularly in combination with Differential Privacy (I10, I13, I25). Some, however, were not so convinced by the privacy-preserving capabilities of synthetic data since it has no \textit{\say{provable guarantees}} (I10) and is generated \textit{\say{from the original data, which doesn't really solve the privacy problem}} (I7). Furthermore, I20 and I21 note that synthetic data may still leak sensitive characteristics of the original data. Common challenges were also cited, such as lack of variability, difficulty in controlling synthetic data generation, and quality of models trained on synthetic data.


\textbf{Private Training.}
Although several participants commented on Augmented Training Data (4/25), this was often in the context of model training, and no connections were made to privacy specifically, such as with adversarial training. 

Similarly, Model Diversification (5/25) was familiar to some for model robustness and improving performance, but not for privacy: \textit{\say{I've never thought about it in terms of privacy improving.}} (I5). Only I24 referenced this technique as a privacy risk mitigation, although with skepticism.

Federated Learning (9/25) was a familiar concept, but few had practical experiences. Multiple participants (5/25) pointed out challenges in practical applicability, particularly with defining the threat model. I7 advocated for Federated Learning as an efficiency improvement rather than a privacy risk mitigation strategy, while I21 supported its use to comply with European regulations, especially with medical data.

\textbf{Post Hoc Privatization.}
Model Explainability (7/25) was perceived as promising, yet still remaining very much in research. Explainable models were tied to privacy in the way that they increase transparency on \textit{\say{what type of information is actually stored}} (I23). Nevertheless, I7 and I16 doubted the feasibility of explainability given current techniques.  

Interface-side Mitigations (7/25) were largely interpreted to be guardrails implemented in AI systems, seen to be very relevant by the participants familiar with them. The main challenge, though, becomes how to develop guardrails in unison with rapidly evolving technologies: \textit{\say{This technology is growing so fast that guardrails are not catching up yet.}} (I8). Deploying guardrails at scale was named as a challenge (I22), and the main limitation remains that \textit{\say{no matter how many guardrails you make, there will always be cases where you will have issues, where you won't be able to guard well}} (I23). 

While Machine Unlearning (7/25) was agreed upon to be interesting and potentially useful, several participants (3/25) noted its complexity to implement and its lack of reliability in its current form. Furthermore, I1 and I7 both provide the insight that if the need for unlearning arises, \textit{\say{the damage is already done}} (I1), making it a \textit{\say{last resort}} option (I7). 

Two participants mentioned Private Inference (2/25) as a \textit{\say{necessity in all deployed systems}} (I7) and \textit{\say{ironclad solution by design}} (I22); few insights beyond this were gained.

\begin{tcolorbox}[boxsep=3pt,left=1pt,right=1pt,top=1pt,bottom=1pt]
\textbf{Key Finding:} AI privacy risk mitigation strategies received varying degrees of attention in the interviews, with popular choices such as \textit{Differential Privacy} and \textit{Synthetic Data}, alongside lesser-known mitigations such as \textit{Functional Secret Sharing}.
\end{tcolorbox}


\subsubsection{Other mitigation strategies}
Our participants brought new mitigation strategies to light:

\begin{itemize}
    \itemsep 0em
    \item \textit{Organizational Measures}: Multiple participants (6/25) mentioned organizational measures as effective, despite serving in a technical role. Examples include trainings, guidelines, and strong data governance. 
    \item \textit{Output Sanitization}: I11 mentions output sanitization, particularly in the form of hallucination control and ensuring that no PII is contained in the outputs of models.
    \item \textit{Monitoring}: I18 and I19 were proponents of continuous monitoring, to be used in conjunction with guardrails.
    \item \textit{Model-specific Mitigations}: Two participants mentioned model-specific mitigations, such as privacy-conscious LLMs or smaller, specialized models that reduce the risks brought about by large, data-hungry models.
    \item \textit{Auditing}: I.e., stricter audits for AI providers: \textit{\say{We have audits for finance, right? (...) We don't have audits for AI (...) So we need to move towards where we actually audit all these companies for risk and safety.}} (I23).
\end{itemize}

\subsection{The Future of Privacy in AI}
\label{sec:future}
Finally, we explored with participants what stakeholders should be involved in safeguarding AI privacy as well as socio-technical opportunities for improving AI privacy.

\subsubsection{Key stakeholders for safeguarding AI privacy}
We asked the participants which entity is ultimately responsible for safeguarding privacy in AI.

\begin{itemize}
    \itemsep 0em
    \item \textbf{The Government and Regulatory Bodies (9/25)}: these authorities can best protect privacy through laws and regulations. In addition to \say{restricting}, it was suggested that governments should also \textit{\say{open opportunities}} (I21), such as to learn about the responsible use of AI. 
    \item \textbf{AI Providers (8/25)} as the organizations \textit{\say{holding the key}} (I16) to AI, these entities have a responsibility to increase the privacy, transparency, and trust in their models, as well as to give users better control of their privacy.
    \item \textbf{End Users (4/25)}: some developers suggested that the responsible use of AI is \textit{\say{ultimately, the responsibility of the user}} (I17), and the pressure to preserve privacy must also come from end users, since \textit{\say{if the customer is going to pay only for a privacy-friendly product, then there is no other choice for the companies.}} (I25).
    \item \textbf{Academia (4/25)}: the role of universities and other educational institutions was seen as important to \textit{\say{reach the masses (...) to explain things in a balanced and objective way}} (I14), as well as to continue advancing privacy research efforts, with a focus on practical applications.
\end{itemize}

\subsubsection{Opportunities for improving AI privacy}
Some participants were not very optimistic about addressing or improving the privacy concerns around AI:
\begin{quote}
    \textit{\say{We are not future-ready. I don't know if anyone is future-ready, and or even the short future-ready. We are just trying to save the day.}} (I20)
\end{quote}
However, many participants shared ideas on what would help reinforce privacy-conscious, responsible AI in the future. They also expressed clear needs that would help them feel better equipped to mitigate AI privacy risks going forward.

\textbf{Regulations, Standards, and Guidelines (14/25).}
Many participants hoped for a future with higher regulation of training data, more transparency, and increased safety checks when AI models are released. 
Especially in enterprise contexts, legal compliance was seen as the \textit{\say{highest concern}} (I11).
Furthermore, some participants criticized current regulations for not being well-equipped to handle AI cases, and that privacy rights might not be protected well (I24).

Among the main expectations regarding privacy in AI is the development of clearer, stricter regulations – an \say{AI equivalent of GDPR} (I5). 
Similarly, standardization would be an advancement in making it \textit{\say{an official profession to be able to work with these models}} (I18). It was emphasized, however, that it is important that regulations do \textit{\say{not hinder innovation itself, but hinder malicious use of it}} (I2), and that \textit{\say{bad guys do not follow regulations, but they are free to use AI}} (I17), reinforcing the belief that regardless of the regulations, malicious actors will find a way to circumnavigate safeguards: 
\begin{quote}
    \textit{\say{There will always be bad actors. If you try to regulate them, it won't work. (...) Your best option is to regulate big companies. That's the only thing you can do.}} (I23)
\end{quote}

In terms of supporting AI developers in mitigating AI privacy risks in their work, a wish for official regulatory guidance was expressed, since as I7 put it, \textit{\say{what's missing is clear direction and clear advice, what is okay, what is not okay.}} (I7)

\textbf{Tools and Testing Systems (11/25).}
Many participants pointed to the lack of tools available to detect privacy risks, as well as to implement and evaluate privacy mitigations. Privacy threat modeling was explicitly mentioned twice (I13, I24). Notably, it was mentioned that misuse is important to address, but monitoring and detecting it is difficult.


\textbf{Education and Awareness (10/25).}
The need for better education on privacy in AI was advocated for, particularly among end users, developers, and legal professionals. Many participants supported the idea that once awareness of privacy risks is made clear, then more informed action can be taken. 

This was particularly tied to the discussion of ethical aspects of AI privacy risks, which was seen to be exacerbated by the current \textit{\say{wild west}} (I19) in AI.
In this, developers highlighted the point that teaching responsible AI is paramount. 

\begin{quote}
    \textit{\say{People who deal with it every day need to explain it in simple terms, not necessarily technical terms. (...) You have to lower the hurdle.}} (I14)
\end{quote}

\textbf{Transparency (6/25).}
Many participants called for higher levels of transparency in the AI field, particularly from the creators of AI systems. This could entail clear documentation on the data processing, steps taken to train models, as well as interface-side privacy disclaimers in simple language. 


\textbf{Interdisciplinary Collaboration (5/25).}
Other participants supported collaboration between practitioners of varying backgrounds, both legal and technical, to help build up competencies and reduce miscommunication. This would help to \textit{\say{get support from different teams or different other groups which work with specific things in AI such as privacy}} (I6). Another example given was the sharing of experience reports.

\begin{quote}
    \textit{\say{I think lots of bad decisions that happen along the way are due to some form of miscommunication or things that happen between people with different backgrounds who are maybe not able to fully understand each other.}} (I3)
\end{quote}

\begin{tcolorbox}[boxsep=3pt,left=1pt,right=1pt,top=1pt,bottom=1pt]
\textbf{Key Finding:} The interviewed developers indicated a number of key stakeholders in AI privacy, particularly the government and AI providers. Together, these stakeholders are tasked with providing clearer guidelines and more usable mitigation tools, with the goal of increased awareness, transparency, and collaboration. 
\end{tcolorbox}

\section{Discussion}
We reflect on our findings, re-examining the lessons of our results in the context of our original research questions.

\subsubsection*{RQ1: Navigating the Sea of Privacy Risks}
Our discussions with 25 AI developers reveal that there is surprisingly little agreement within the technical community over which AI privacy risks are the most pressing. In fact, out of the 25 ranking exercise responses, we observed 23 unique orderings, supporting that there is indeed a clear lack of consensus. Interestingly, the risks of data management, data memorization, and misuse ranked amongst the top 3, each with seven \#1 votes, yet they also appeared as last in the ranking a number of times (4, 4, and 5, respectively). This implies that although certain risks are generally perceived as more urgent than others, this is not uniformly the case. 

The variety in responses raises potential concerns. The notion that ethical considerations currently trail behind the proliferation of AI was expressed often, 
adding to a seeming confusion over what risks are most important. Additionally, we observe little cohesion between rankings even when isolating attributes such as educational level, experience, or company size. This points to an industry-wide lack of unification on privacy risk prevalence, displaying that companies may scope privacy risks quite differently or, rather, may not be scoping privacy risks at all. Although this is natural given the variety of business contexts and the relative nascence of AI privacy risk knowledge, these findings call for a more comprehensive understanding among developers and companies, specifically on how to prioritize privacy risk assessments and how one may differentiate between different risk classes.

We learn that assessing privacy risks is currently very personal -- reasons behind a specific ranking are often attributed to an internal hierarchy influenced by some self-guided reasoning pattern, which can be especially influenced by a developer's specific work context. This resonates with previous findings on general privacy work among developers \cite{10.1145/3589951,lee2024don}, and suggests a lack of clear guidelines from companies or authorities, which forces developers to navigate the sea of privacy risks on their own. Moreover, we find that in lieu of clear guidelines, developers often fall back to internal reasoning patterns on how to assess and scope privacy risks. However, such reasoning may become clouded by confounding factors, such as the uncertainty between privacy risks and model hallucinations (I18). While this suggests the need for more guidance from researchers and authorities, it also points to one potential benefit: allowing developers to explore novel mitigation techniques and build up best practices.

Given this current state, regulators looking to the industry for best practices on AI privacy risk mitigation may be left without clear answers. Thus, attempts to regulate AI without validated industry standards and accepted mitigation strategies may lead to premature regulations that force AI developers into non-optimal solutions, thereby creating more friction.

\subsubsection*{RQ2: Mitigations Abound, but Adoption is Lagging}
In discussing current mitigation strategies for AI privacy risks, we learned that general awareness of the 19 presented techniques is high, and beyond this, developers shared a number of other strategies they believed to be mitigations. Certain mitigation strategies received considerable attention, such as Synthetic Data and Differential Privacy, while others garnered nearly no comments (e.g., Functional Secret Sharing). Participants also mentioned mitigations not popularly covered in the literature, such as active monitoring for quality assurance. These findings show a development in the conclusions drawn by Boenisch et al. \cite{10.1145/3473856.3473869}, suggesting that developer awareness of privacy mitigations has increased in the past few years.

We observed that some strategies received quite a deal of attention, such as Alignment, Synthetic Data, and Differential Privacy. This attention, one can argue, echoes current research and media attraction to such topics, implying that mitigation awareness may coincide closely with prevalence in academic literature or popular media. This points to the reach of academic publications, as well as exhibits the degree to which developers may relate to popular media topics.

Much of the commentary on mitigations also sheds light on the \say{go-tos} in the industry, such as the blanket terms of anonymization and data cleaning, which were largely met with approval and reported to be used often.
In contrast, many of the \say{advanced} techniques, including those in Data Augmentation, Private Training, and Post Hoc Privatization, were often perceived to have significant shortcomings, which hinder practical adoption, such as high costs, slow speeds, and/or loss of utility or functionality, a finding that echoes previous work \cite{10538656,iceis24}. Thus, a clear divide was formed between the widely accepted, yet less advanced mitigations and those that show high promise but low widespread practicality.

The implications of this divide become significant when addressing the practical adoptability of advanced PETs such as Differential Privacy, where it will become important to distinguish clear practical limitations and resistance to change from more familiar \say{anonymization techniques} \cite{garrido2023lessons}. Similarly, the case of Synthetic Data teaches us that wide application does not equate to sound privacy protection, as the promise of this technique was met with equally as many concerns over its ability to provide actual privacy guarantees. This comes into contrast with other advanced techniques, which achieved little recognition from any of the interviewed developers. Furthermore, the question of familiarity may be influenced by the potential \textit{use cases} of a mitigation strategy, where a lesser-known strategy may be linked to its limited applicability to specific or niche AI use cases, e.g., with machine unlearning.

As such, we see it as crucial to raise awareness of the wide range of privacy risk mitigation strategies to more developers. Beyond this, it is important that the burden of learning about new mitigations is not placed solely on individual developers, but rather the discussion should involve more people, whether through the establishment of community norms or fostering interdisciplinary collaboration. In addition, regulators and other supervisory authorities, such as standards organizations, could work to provide guidance and recommendations on the use of state-of-the-art mitigations to support privacy-conscious AI development. These steps can help developers reach a better consensus on how to reason about privacy risks in AI.

\subsubsection*{RQ3: Becoming Future-ready}
In the interviews, we received various recommendations on how to advance understanding of privacy risks and adoption of mitigation strategies. These include clear calls to action for key stakeholders involved in the AI pipeline, from researchers and developers, to enterprise leadership and end users.

Many of the interviewed developers believed that AI providers, or more generally those with the resources to develop, train, and deploy large-scale AI systems, are the ones that \say{hold the keys} to developing more privacy-conscious AI. This, however, will not be done without a clear demand from enterprise customers and end users to make privacy a priority, thereby incentivizing privacy to ensure the continued success of an AI product or service; this, however, was seen as the \say{happy path}, or more optimistic view of the future.

As noted by participants, privacy-conscious demand from users may be hindered by their focus on functionality over privacy, and a greater privacy awareness among users will not come without intervention. Similarly, in the enterprise context, a lack of awareness of the privacy risks when entrusting company data in the hands of AI providers may pose a risk to further adoption. Raising awareness thus becomes the task of researchers and educators, where university curricula and other programs should prioritize privacy and other ethical considerations in AI education, similar to calls emphasizing privacy education when teaching software development~\cite{prybylo2024evaluating}.

The final point that reached consensus is that governments and other supervisory authorities can directly influence ethical AI practices within companies by enforcing strict regulations and providing official guidelines that promote the use of advanced PETs. This influential role is emphasized by previous work \cite{iceis24}, requiring collaboration with researchers for guidance on technological solutions for privacy risk mitigation.

In this, we uncover the blueprints for an ecosystem in which all parties -- AI developers, end users, researchers, and regulators -- must actively participate, where privacy is simultaneously incentivized and enforced. Only with these forces working in parallel will there be an advancement towards a future where AI privacy risks are well-defined, mitigations are not only promising but also offer an attractive privacy-utility trade-off, and developers and users alike feel equipped to contribute to a more privacy-friendly future of AI. 

\section{Conclusion}
We interviewed 25 AI developers to understand their perceptions of known privacy risks in AI, as well as their familiarity with current strategies to mitigate these risks. In exploring the perceived importance of these risks, we learned that there exists a multitude of reasoning patterns as to why one risk may be more urgent than another. We also observed that while general familiarity with mitigation strategies is strong, few developers in our study could point to concrete cases in which more advanced privacy techniques (e.g., Differential Privacy) are currently employed. Developers shared that building more privacy-conscious AI systems necessitates the involvement of large AI providers, end users, researchers, educational institutions, and governmental bodies. As a whole, our findings highlight the need for better awareness of privacy risks, improved privacy-enhancing tooling in AI, and greater cooperation in order to improve AI privacy for users and enterprises.

\section*{Acknowledgments}
This research is based on work that has been partially funded by Google. Any opinions, findings, conclusions, or recommendations expressed in this work are those of the authors and do not necessarily reflect the views of Google.


{
\bibliographystyle{plain}
\setlength{\bibsep}{6.4pt}
\bibliography{usenix2025_SOUPS}
}

\onecolumn
\appendix

\section{AI Privacy Risks}
\label{sec:appendix:risks}

\begin{table}[h]
\centering
\scriptsize
\begin{tabular}{|p{\linewidth}|}
\hline
\textbf{(R1) Data Management} \cite{shahriar2023survey}\\
As the training and fine-tuning of AI requires massive amounts of data to be accurate and robust, large amounts of data must be collected and stored, posing privacy risks to those from whom the data originated.\\
\textbf{(R2) Data Memorization and Leakage} \cite{bsi2024ai,rahman2023security, shahriar2023survey, smith2023identifying, vassilev2024adversarial}  \\
Larger models have the tendency to “memorize” some of their training data, particularly the data that is more unique or rare occurring. Data leakage occurs when models unintentionally expose their memorized data, and in the case of prompting, malicious users may be able to reconstruct or extract data with careful prompting. \\
\textbf{(R3) Unintended Downstream Usage (of Sensitive Prompts and User Metadata)} \cite{bsi2024ai,yan2024protecting}\\
User inputs and queries may be used for originally unintended purposes, for example, building up user profiles based upon contextual information, or sensitive company data being captured and used downstream for model fine-tuning. \\
\textbf{(R4) Adversarial Inference and Inversion} \cite{rahman2023security,rigaki2023survey, shahriar2023survey, smith2023identifying,vassilev2024adversarial, yan2024protecting, jayaraman2022attribute, bsi2024ai, YAO2024100211} \\
A class of attacks against AI referred to as inference attacks include an adversary trying to distinguish whether a particular instance (member) was in the training data, attempting to recover attributes of the training data, making broader inferences about the underlying data, or reverse-engineering model-specific parameters.\\
\textbf{(R5) Misuse via Harmful Applications} \cite{council2023ghost,golda2024privacy,mcgraw2024architectural,YAO2024100211} \\
The predictive and generative power of AI opens the door for misuse in downstream applications, where malicious users leverage AI for nefarious purposes (e.g., deepfakes). \\
\hline
\end{tabular}
\caption{AI privacy risks identified through the literature review, as introduced to the participants.}
\label{tab:risks}
\end{table}

\section{Mitigations for AI Privacy Risks}
\label{sec:appendix_mitigations}

\begin{table}[ht!]
\centering
\scriptsize
\begin{tabular}{|p{\linewidth}|}
\hline
\multicolumn{1}{|c|}{\textbf{Organizational}} \\ \hline
\textbf{(M1.1) Management of (Training) Data} \cite{bsi2024ai, golda2024privacy}\\
Mitigating privacy risks begins with the responsible handling of training data, for example in trusted and secured data warehouses, as well as proper data governance structures.\\
\textbf{(M1.2) Alignment} \cite{bsi2024ai}\\
An important organizational measure involves the aligning of AI to human values, a step that can help to ensure trustworthiness and reduce the risk of adversaries compromising models. \\
\textbf{(M1.3) Trusted Execution Environments} \cite{curzon2021privacy, yan2024protecting} \\
Secured hardware environments where data processed within cannot be read or tampered with by outside parties or code. \\ \hline
\multicolumn{1}{|c|}{\textbf{Data Preprocessing}} \\ \hline
\textbf{(M2.1) Input Sanitization/Data Cleaning} \cite{curzon2021privacy,bsi2024ai,rahman2023security, smith2023identifying, yan2024protecting,YAO2024100211}\\
When preparing data for model training, a wise preprocessing step includes data cleaning, in which explicit sensitive information is removed, such as phone numbers. \\
\textbf{(M2.2) Anonymization/Pseudonymization} \cite{bsi2024ai}\\
Similarly, explicit personally identifiable information (PII) can be removed via anonymization or pseudonymization techniques. \\
\textbf{(M2.3) Outlier Detection} \cite{mcgraw2024architectural,shahriar2023survey,vassilev2024adversarial,YAO2024100211}\\
Outlier detection methods aim to identify and detect outliers in the training data, as these datapoints might be more readily memorized.  \\ \hline
\multicolumn{1}{|c|}{\textbf{Data Augmentation}} \\ \hline
\textbf{(M3.1) Differential Privacy} \cite{curzon2021privacy,bsi2024ai,golda2024privacy,smith2023identifying,vassilev2024adversarial, yan2024protecting,YAO2024100211}\\
Differential Privacy is a mathematically grounded notion of privacy that usually involves adding random noise to query outputs to inject plausible deniability into computations on potentially private data (attributes).  \\
\textbf{(M3.2) Group Privacy} \cite{council2023ghost,curzon2021privacy,vassilev2024adversarial} \\
The notions of k-anonymity, l-diversity,   and t-closeness concern themselves with group privacy, where the goal is to provide a certain level of indistinguishability between members of a dataset. \\
\textbf{(M3.3) Data Transformation} \cite{curzon2021privacy,shahriar2023survey}\\
Storing data in “transformed” forms, such as in embedding format or in reduced dimensions, may help to obfuscate raw sensitive data. \\
\textbf{(M3.4) Homomorphic Encryption} \cite{golda2024privacy,yan2024protecting}\\
Advanced privacy-preserving techniques based on encryption can be leveraged such that computation on raw user data is not necessary. \\
\textbf{(M3.5) Functional Secret Sharing} \cite{yan2024protecting}\\
In some privacy-preserving mechanisms, user data is “split” among a pool of users, such that no single data point must be shared in full. \\
\textbf{(M3.6) Synthetic Data} \cite{curzon2021privacy}\\
Rather than use the original training data, some training strategies may opt to use synthetically generated data,   which ideally shares a similar distribution of the original data. \\ \hline
\multicolumn{1}{|c|}{\textbf{Private Training}} \\ \hline
\textbf{(M4.1) Augmented Training Data} \cite{curzon2021privacy,bsi2024ai,rahman2023security}\\
Adding augmented samples to training data can help to improve model robustness, thereby reducing the risk of privacy breaches. \\
\textbf{(M4.2) Model Diversification} \cite{rahman2023security}\\
Some training procedures may opt to train a variety of models (or parameters) to improve robustness in decision-making and to mitigate any vulnerabilities of one single model. \\
\textbf{(M4.3) Federated Learning} \cite{golda2024privacy,rahman2023security, yan2024protecting}\\
The training of models is performed locally in a distributed fashion, where only model updates are shared with a central aggregator. \\ \hline
\multicolumn{1}{|c|}{\textbf{Post Hoc Privatization}} \\ \hline
\textbf{(M5.1) Model Explainability} \cite{bsi2024ai,golda2024privacy,rahman2023security, shahriar2023survey}\\
Focusing on the development of explainable models can detect model vulnerabilities, as well as verify the integrity of a model. \\
\textbf{(M5.2) Interface-side Mitigations} \cite{bsi2024ai,smith2023identifying,vassilev2024adversarial}\\
Safeguarding the interface between users and models includes detecting suspicious queries, limiting the number of queries, and building guardrails. \\
\textbf{(M5.3) Machine Unlearning} \cite{bsi2024ai,smith2023identifying, yan2024protecting} \\
The ability to remove a single user's contributions to a training data set upon request, without the need for complete retraining. \\
\textbf{(M5.4) Private Inference} \cite{bsi2024ai}\\
Augmenting model inference (computation on unseen data) with privacy-preserving solutions such as Differential Privacy or Homomorphic Encryption adds an extra layer of privacy protection for users inputting potentially sensitive data. \\ \hline
\end{tabular}
\caption{Mitigations for AI privacy risks identified through the literature review, as introduced to the participants.}
\label{tab:mitigations}
\end{table}

\newpage

\section{Recruitment Message}
The following message was used to contact and recruit the interview participants described in this work. Where applicable, we personalized this message to tailor to a specific interview candidate.

\par\noindent\rule{\textwidth}{0.4pt}

Subject: \textbf{Privacy Risks in General-Purpose AI} \newline

Dear [TITLE][NAME], \newline

I hope this message finds you well. \newline

[One-line introduction of the message author.] \newline


We are currently conducting a study investigating the practical perspectives on Privacy Risks in General-Purpose AI Systems. For this, we are searching for technical experts in the field of AI to share their opinions on the topic. \newline

Based on your impressive professional background and your current role as [ROLE], I believe you could bring very valuable insights to our study and would like to invite you to participate in a 1-hour Zoom interview at a time that is convenient for you. \newline

I would really appreciate it if you could participate, and I look forward to hearing from you! \newline

Best regards, \newline

[Email-style signature.]

\par\noindent\rule{\textwidth}{0.4pt}

\section{Interview Participant Demographics}
\label{sec:appendix_interviewees}

\begin{table*}[hbtp]
\centering
\footnotesize
\resizebox{\linewidth}{!}{
\begin{tabular}{l|l|l|l|l|l|l|l}
\toprule
\textbf{ID} & \textbf{Position} & \textbf{Education} & \textbf{Industry Domain} & \textbf{Org. size} & \textbf{Country}  & \textbf{Exp. (years)} & \textbf{AI-centered} \\
\midrule
I1 & Senior Researcher & Doctorate & Information Technology & Large & Germany & 3 - 5 & No \\
I2 & Senior Researcher & Doctorate & Information Technology & Large & Germany & 5 - 10  & No \\
I3 & Generative AI Engineer & Doctorate & Information Technology & Small & Sweden & 3 - 5  & Yes \\
I4 & Senior AI Developer & Doctorate & Information Technology & Small & Spain & 5 - 10  & Yes \\
I5 & Software Architect & Doctorate & Information Technology & Large & Germany & 10 - 20  & No \\
I6 & AI Engineer & Master's & Information Technology & Medium & Germany & 1 - 3  & Yes \\
I7 & CEO & Doctorate & Information Technology & Small & Germany & 10 - 20  & Yes \\
I8 & MLOps Engineer & Master's & Information Technology & Medium & Ukraine & 5 - 10 & No \\
I9 & Managing Director & Master's & Finance & Large & Germany & 20+  & No \\
I10 & Senior Researcher & Doctorate & Information Technology & Large & Germany & 10 - 20  & No \\
I11 & Head of AI \& Advanced Analytics & Master's & High tech, Telecom & Large & \text{India} & 10 - 20 & Yes \\
I12 & Head of the Innovation Department & Bachelor's & Finance & Large & Germany & 20+  & No \\
I13 & Senior Privacy Engineer & Master's & Advertising & Small & Italy & 3 - 5 & Yes \\
I14 & Deputy Head of Department & Master's & Information Technology & Large & Germany & 10 - 20 & No \\
I15 & Principal AI Consultant & Doctorate & Information Technology & Medium & Croatia & 10 - 20 &  No \\
I16 & Senior AI Scientist & Doctorate & Information Technology & Large & Sweden & 10 - 20  & Yes \\
I17 & AI Engineer and Researcher & Master's & Electronics & Medium & Ukraine & 10 - 20 & No \\
I18 & Founder / ML engineer & Doctorate & Information Technology & Micro & Netherlands & 10 - 20 & Yes \\
I19 & Lead Data \& AI Consultant & Doctorate & AI Consulting & Small & Switzerland & 20+  & Yes \\
I20 & Head of AI Division & Doctorate & Information Technology & Small & Germany & 20+  & Yes \\
I21 & AI Engineer & Master's & Information Technology & Medium & United Kingdom & 3 - 5 & Yes \\
I22 & AI Engineer & Master's & Information Technology & Small & France & 20+ & Yes \\
I23 & Senior AI Engineer & Master's & Information Technology & Micro & Belgium & 3 - 5 & Yes \\
I24 & Senior Privacy Engineer & Master's & Information Technology & Large & Austria & 5 - 10 & Yes \\
I25 & Senior Privacy Engineer & Doctorate & Automotive & Large & Germany & 5 - 10  & No\\
\bottomrule
\end{tabular}
}
\caption{Interview participant demographics. \textit{AI-centered} denotes whether AI is integral to the main product or service of the participant's company.}
\label{tab:interview_participants}
\end{table*}

\newpage

\section{Interview Guide}
\label{sec:interview_guide}
\small
\textbf{Disclaimer} \newline
By taking part in this interview, you agree for the audio of the interview to be recorded and transcribed for further analysis. The interview transcripts will not be shared with anyone outside of our immediate research team. You also agree for direct quotes from the interview to be used for publication, and that such quotes will be attributed in a pseudonymized form. No PII or any other personal data will be shared or attributed. Please confirm your consent to these terms.
\vspace{10pt}

\textbf{Definitions} \newline
\underline{General Purpose AI System (GPAIS)}: an AI system that is able to perform generally applicable functions such as image/speech recognition, audio/video generation, pattern detection, question answering, translation, etc., and is able to have multiple intended and unintended purposes. Within the scope of GPAIS, we investigate Large Language Models, Diffusion Models, and Multimodal Models.

\textbf{AI Background}
\begin{enumerate}
    \item[1.] *With which of the abovementioned model types do you have (the most) experience?
    \item[2.] Do you use AI systems in your role at work?
    \begin{itemize}
        \item If yes, can you describe in what capacity?
        \item Have you trained or fine-tuned AI models?
        \item Which models have you used?
        \item What applications have you integrated AI into?
    \end{itemize}
    \item[3.] How central is AI to the product you work on?
\end{enumerate}

\textbf{Risk Awareness + Mitigation Strategies} 

\begin{enumerate}
\item[4.] *What do you think are some of the privacy risks of General-Purpose AI?
\item[5.] Are these risks general privacy risks, or specific to GPAIS? Are any risks amplified by GPAIS?
\item[6.] [Ordering Exercise] \newline
\textit{In your opinion, which of these poses the greatest risk?}
\item[7.] How would you rate the importance of other non-technical privacy risks, in particular legal/regulatory and ethical/societal risks?
\item[8.] Can you think of any other privacy risks in GPAIS that we have not covered in the previous question?
\item[9.] *Are you familiar with any mitigation strategies for the risks we have just covered?
\begin{itemize}
    \item Have you implemented any of them?
    \begin{itemize}
        \item If yes, can you describe your experience with this? Were there any challenges?
        \item If not, what were the blockers?
    \end{itemize}
    \item Who normally implements these measures? At which stage? [Development, Deployment, Application]
\end{itemize}
\item[10.] Are you aware of any other mitigation strategies that could be employed?
\item[11.] Out of the mitigation strategies mentioned, which do you believe to be most effective?
\end{enumerate}

\textbf{Privacy Risk Mitigation Readiness}

\begin{enumerate}
    \item *In your opinion, which privacy risks do you feel can currently be best mitigated given current mitigation strategies?
    \begin{enumerate}
        \item Which risks are more difficult or not currently possible to mitigate?
    \end{enumerate}
    \item *Which of these risks do you personally feel best equipped to address? Why?
    \item What would help you feel more equipped?
    \item How do you learn about new technologies for privacy risk mitigation?
\end{enumerate}

\textbf{Looking Forward}

\begin{enumerate}
    \item *What do you envision the next few years will look like with regard to privacy and General-Purpose AI?
    \begin{enumerate}
        \item Do you think privacy will become a bigger/smaller issue in the near future with respect to the continued development of General-Purpose AI?
        \item How do you think these issues will be handled? (reduced adoption, increased mitigation, etc.)
        \item What would make addressing these issues easier/harder?
        \item What would make people more confident in AI systems with respect to privacy? Who should be responsible for this?
    \end{enumerate}
\end{enumerate}

\textbf{Other}

\begin{enumerate}
    \item Is there any other aspect of this topic we may have missed but you feel is important to discuss?
    \item Can you refer anyone who would also be able to contribute to this discussion?
\end{enumerate}

\newpage

\section{Risk Ranking Survey}
\label{sec:risk_survey}

\textbf{Privacy Risks of General-Purpose AI Systems}

\begin{enumerate}
    \item \textbf{Data Management} \newline
    As the training and fine-tuning of GPAIS requires massive amounts of data to be accurate and robust, large amounts of data must be collected and stored, posing privacy risks to those from whom the data originated.
    \item \textbf{Data Memorization and Leakage} \newline
    Larger models have the tendency to “memorize” some of their training data, particularly the data that is more unique or rare occurring. Data leakage occurs when models unintentionally expose their memorized data, and in the case of prompting, malicious users may be able to reconstruct or extract data with careful prompting.
    \item \textbf{Unintended Downstream Usage of Sensitive Prompts and User Metadata} \newline
    User inputs and queries may be used for originally unintended purposes, for example building up user profiles based upon contextual information, or sensitive company data being captured and used downstream for model fine-tuning.
    \item \textbf{Adversarial Inference and Inversion} \newline
    A class of attacks against GPAIS referred to as inference attacks include an adversary trying to distinguish whether a particular instance (member) was in the training data, attempting to recover attributes of the training data, making broader inferences about the underlying data, or reverse-engineering model-specific parameters.
    \item \textbf{Misuse via Harmful Applications} \newline
    The predictive and generative power of GPAIS opens the door for misuse in downstream applications, where malicious users leverage AI for nefarious purposes (e.g., deepfakes).
\end{enumerate}

\textit{[Sorting Exercise]}

In your opinion, which of these risks is most urgent to address / most harmful?

\begin{itemize}
    \item Data Management
    \item Data Memorization and Leakage
    \item Unintended Downstream Usage of Sensitive Prompts and User Metadata
    \item Adversarial Inference and Inversion
    \item Misuse via Harmful Applications
\end{itemize}

\section{Codebooks}
We make our codebooks public, i.e., resulting from the conducted thematic content analysis on the interview data. We split our analysis results into three codebooks: \textit{Risks}, \textit{Mitigations}, and \textit{Future}, corresponding to findings related to each of our three research questions. The codebooks can be found in PDF form at: \url{https://drive.google.com/drive/folders/1BPTOF_JOIsSPqK4WJ-uVaPXdoiIIcVxk?usp=sharing}

\newpage

\section{Mitigations Survey}
\label{sec:mitigations_survey}
\small
\textbf{Mitigations for Privacy Risks in General-Purpose AI Systems} \newline

Thank you once again for taking part in our interview study! 
We kindly ask you to help us structure some of the feedback we received during the interviews regarding the possible mitigations for the privacy risks of General-Purpose AI systems.
\newline

In particular, we will ask you to assess your familiarity and experience with the mitigations that we presented during the interview, scored on a 4-point scale: \textit{never heard of it}, \textit{heard of it}, \textit{considered using it}, \textit{Used/using it}.

NOTE: Please respond in the way applicable at the time of the interview, i.e., if you never heard of a mitigation before our interview, please select \textit{never heard of it}. In cases of uncertainty, please just pick the most fitting response!
\newline

The remainder of this form consists of five sections that list the different mitigations we discussed, grouped into five categories: \textit{Organizational}, \textit{Data Preprocessing}, \textit{Data Augmentation}, \textit{Private Training,} and \textit{Post Hoc Privatization}. 
\newline

The survey should take no longer than 5 minutes to complete.
\newline

Before proceeding, please indicate your name - this is just for us to keep track of who submitted the form, of course, these results will also be pseudonymized!
\newline

Thank you very much!

\begin{enumerate}
    \item[1.] *Your name
    \end{enumerate}

\underline{\textbf{Organizational}}

\begin{itemize}
    \item \textbf{Management of (Training) Data}: Mitigating privacy risks begins with the responsible handling of training data, for example in trusted and secured data warehouses, as well as proper data governance structures.
    \item \textbf{Alignment}: An important organizational measure involves the \textit{aligning} of GPAIS to human values, a step that can help to ensure trustworthiness and reduce the risk of adversaries compromising models.
    \item \textbf{Trusted Execution Environments}: Secured hardware environments where data processed within cannot be read or tampered with by outside parties or code.
\end{itemize}

\begin{enumerate}
    \item[2.] Please assess your familiarity with the above privacy risk mitigation approaches: \newline
\textit{[Response options: {\textit{never heard of it}, \textit{heard of it}, \textit{considered using it}, \textit{Used/using it}}.]}
\begin{enumerate}
    \item *Management of (Training) Data
    \item *Alignment
    \item *Trusted Execution Environments
    \end{enumerate}
\end{enumerate}

\underline{\textbf{Data Preprocessing}}

\begin{itemize}
    \item \textbf{Input Sanitization/Data Cleaning}: When preparing data for model training, a wise preprocessing step includes data cleaning, in which explicit sensitive information is removed, such as phone numbers.
    \item \textbf{Anonymization/Pseudonymization}: Similarly, explicit personally identifiable information (PII) can be removed via anonymization or pseudonymization techniques.
    \item \textbf{Outlier Detection}: Outlier detection methods aim to identify and detect outliers in the training data, as these datapoints might be more readily memorized.
\end{itemize}

\begin{enumerate}
    \item[3.] Please assess your familiarity with the above privacy risk mitigation approaches: \newline
    \textit{[Response options: {\textit{never heard of it}, \textit{heard of it}, \textit{considered using it}, \textit{Used/using it}}.]}
\begin{enumerate}
    \item *Input Sanitization/Data Cleaning
    \item *Anonymization/Pseudonymization
    \item *Outlier Detection
    \end{enumerate}
\end{enumerate}

\underline{\textbf{Data Augmentation}}

\begin{itemize}
    \item \textbf{Differential Privacy (DP) (Randomized Response)}: DP is a mathematically grounded notion of privacy which usually involves adding random noise to query outputs to inject plausible deniability into computations on potentially private data (attributes).
    \item \textbf{Group Privacy}: The notions of k-anonymity, l-diversity, and t-closeness concern themselves with group privacy, where the goal is to provide a certain level of indistinguishability between members of a dataset.
    \item \textbf{Data Transformation}: Storing data in “transformed” forms, such as in embedding format or in reduced dimensions, may help to obfuscate raw sensitive data.
    \item \textbf{Privacy-Preserving Data Aggregation / Homomorphic Encryption}: Advanced privacy-preserving techniques based on encryption can be leveraged such that computation on raw user data is not necessary.
    \item \textbf{Functional Secret Sharing}: In some privacy-preserving mechanisms, user data is “split” among a pool of users, such that no single data point must be shared in full.
    \item \textbf{Synthetic Data}: Rather than use the original training data, some training strategies may opt to use \textit{synthetically generated} data, which ideally shares a similar distribution of the original data.
\end{itemize}

\begin{enumerate}
    \item[4.] Please assess your familiarity with the above privacy risk mitigation approaches: \newline
    \textit{[Response options: {\textit{never heard of it}, \textit{heard of it}, \textit{considered using it}, \textit{Used/using it}}.]}
\begin{enumerate}
    \item *Differential Privacy
    \item *Group Privacy
    \item *Data Transformation
    \item *Privacy-Preserving Data Aggregation / Homomorphic Encryption
    \item *Functional Secret Sharing
    \item *Synthetic Data
    \end{enumerate}
\end{enumerate}

\underline{\textbf{Private Training}}

\begin{itemize}
    \item \textbf{Augmented Training Data}: Adding augmented samples to training data can help to improve model robustness, thereby reducing the risk of privacy breaches.
    \item \textbf{Model Diversification}: Some training procedures may opt to train a variety of models (or parameters) to improve robustness in decision-making and to mitigate any vulnerabilities of one single model.
    \item \textbf{Federated Learning}: The training of models is performed locally in a \textit{distributed} fashion, where only model updates are shared with a central aggregator.
\end{itemize}

\begin{enumerate}
    \item[5.] Please assess your familiarity with the above privacy risk mitigation approaches: \newline
    \textit{[Response options: {\textit{never heard of it}, \textit{heard of it}, \textit{considered using it}, \textit{Used/using it}}.]}
\begin{enumerate}
    \item *Augmented Training Data
    \item *Model Diversification
    \item *Federated Learning
    \end{enumerate}
\end{enumerate}

\underline{\textbf{Post Hoc Privatization}}

\begin{itemize}
    \item \textbf{Model Explainability}: Focusing on the development of \textit{explainable} models can detect model vulnerabilities, as well as verify the integrity of a model.
    \item \textbf{Interface-side Mitigations}: Safeguarding the interface between users and models includes detecting suspicious queries, limiting the number of queries, and building guardrails.
    \item \textbf{Machine Unlearning}: The ability to remove a single user's contributions to a training data set upon request, without the need for complete retraining.
    \item \textbf{Private Inference}: Augmenting model inference (computation on unseen data) with privacy-preserving solutions such as DP or HE adds an extra layer of privacy protection for users inputting potentially sensitive data.
\end{itemize}

\begin{enumerate}
    \item[6.] Please assess your familiarity with the above privacy risk mitigation approaches: \newline
    \textit{[Response options: {\textit{never heard of it}, \textit{heard of it}, \textit{considered using it}, \textit{used/using it}}.]}
\begin{enumerate}
    \item *Model Explainability
    \item *Interface-side Mitigations
    \item *Machine Unlearning
    \item *Private Inference
    \end{enumerate}
\end{enumerate}

\end{document}